\documentclass[aps,reprint,twocolumn,prc,showpacs,superscriptaddress,floatfix]{revtex4-1}
\usepackage{amsmath,amssymb,amsfonts}
\usepackage{graphicx,epsfig}
\usepackage{caption}
\usepackage{xcolor}
\usepackage{times}

\newcommand{\sst}{\scriptscriptstyle}

\begin{document}
\title{Spectral-statistics properties of the experimental and theoretical light baryon and meson spectra}
\author{L. Mu\~noz}
\email{lmunoz@ucm.es}
\affiliation{Grupo de F\'{\i}sica Nuclear, Departamento de F\'{\i}sica At\'omica, Molecular y Nuclear, 
Facultad de Ciencias F\'{\i}sicas, Universidad Complutense de Madrid, 
Avda. Complutense s/n, E-28040 Madrid, Spain}
\author{A. Rela\~no}
\affiliation{Departamento de F\'{\i}sica Aplicada I and GISC,
    Universidad Complutense de Madrid, Av. Complutense s/n, E-28040
    Madrid, Spain}

\begin{abstract}
We compare the statistical fluctuation properties of the baryon and meson
experimental mass spectra with those obtained from theoretical
models (quark models and lattice QCD). We find that for the experimental spectra
the statistical properties are close to those predicted by Random Matrix
Theory for chaotic systems, while for the theoretical ones they are in general
closer to those predicted for integrable systems and safely incompatible
with those of chaotic systems. We stress the importance of the agreement
of the fluctuation properties between experiment and theoretical models,
as they determine the dynamical regime and the complexity of the real
interactions. We emphasize the new statistical method we use, adapted for
properly analyzing the fluctuation properties for very short
spectral sequences.
\end{abstract}


\maketitle

\section{Introduction}

Hadron spectroscopy has played a central role in the study of the strong
interaction helping on its understanding and the development of Quantum
Chromodynamics (QCD). Hadrons constitute bound states for quarks
and gluons, and their accurate description is one of the principal aims of QCD.
However, so far a quantitative and predictive theory of confined states has not
been achieved, hence, in order to study the properties of hadrons we have to
rely on models which have to be consistent with the underlying QCD. Constituent
quark models \cite{Godfrey98} are examples of this kind of modeling.

\underline{\bf Baryons.} It is well known that the number of baryons predicted
by quark models \cite{Capstick86,Bonn01} is substantially larger than what is
observed in meson scattering and production experiments
\cite{Yao06}. This fact raises the problem of {\it missing resonances},
which has opened the door to a huge experimental effort in
recent years to observe and identify these missing states
\cite{missing_exps}. These experiments have to achieve high precision due
to the important background (which can veil resonances)
and the overlap of baryons, as well as the need to survey
different meson production channels and observables. The
procedure to assess the existence of these elusive baryons
consists on analyses of partial waves \cite{Vrana00} and polarization
observables \cite{Dutta02} of the reactions comparing experimental
data from different sources to what is obtained after including
or removing the hypothetical resonance. If data are
better reproduced, the existence of a resonance is possible
but sometimes debatable. The
Particle Data Group (PDG) rates the possible existence of
the resonances based on the quantity and quality of experimental
data. Only after several independent experiments
and analyses, a baryon is awarded a well-established status,
rated with three or four stars.

\underline{\bf Mesons.} In the last decade an enormous experimental effort has
been made in meson spectroscopy with several facilities
conducting research programs \cite{PhysRep} 
 whose main goal has been to find exotic mesons \cite{exotics} which 
do not fit within the quark-antiquark picture of quark models.
This search has been fruitless so far but has put meson
physics at the forefront of scientific research, becoming a thriving
research area with  experimental collaborations in several facilities
---i.e. BES (China) \cite{BES}, 
CLAS at JLab (USA) \cite{CLAS},
COMPASS at CERN (Switzerland) \cite{COMPASS},
J-PARC (Japan) \cite{Jparc} and
Hall D under construction at JLab (USA) \cite{HallD}.

Theoretical research has not been oblivious to this experimental
interest and several quark models of mesons have made their appearance
in the literature \cite{Koll00,Vijande05,Ebert09} trying to
match the low-lying experimental spectrum and complementing the
classic calculation by Godfrey and Isgur \cite{Godfrey85}.  
Among the theoretical developments, it is noteworthy the lattice QCD 
calculation of the meson spectrum 
by the Hadron Spectrum Collaboration (HSC) at JLab \cite{Dudek,Dudek11}, 
although with the drawback of being computed at a high pion mass of 396 MeV.

\underline{\bf Statistics.} As hadrons can be considered as aggregates of quarks
and gluons, the mass spectrum of low-lying baryons or mesons can be understood
as the energy spectrum of an interacting quantum system composed by such quarks
and gluons. Hence, the properties of the masses can be characterized in the
same terms than the energy spectrum of a similar interacting quantum system,
like the atomic nucleus.
Since Wigner discovered that the statistical properties of complex
nuclear spectra are well described by the Gaussian Orthogonal Ensemble (GOE)
of Random Matrix Theory \cite{Porter}, statistical methods have become a
powerful tool to study the energy spectra of quantum systems
\cite{Gomez11,Guhr98}.

Random Matrix Theory allows to establish a connection
between statistical properties of energy spectra and Quantum Chaos. The work
of Berry and Tabor \cite{Berry77}, which shows that integrable systems lead
to energy-level fluctuations that are well described by the Poisson
distribution, and the work of Bohigas, Giannoni and Schmit \cite{Bohigas84},
which conjectured that spectral fluctuation properties of chaotic systems are
well described by Random Matrix Theory (known as the BGS conjecture, later
proved by Heusler {\it et al.} \cite{Heusler07}) can be considered as a
definition of Quantum Chaos in terms of spectral fluctuation
properties. That is, the energy-level fluctuations determine if a system
is chaotic, integrable or intermediate. While for integrable and chaotic
systems these properties are universal, for intermediate systems different
types of transitions from order to chaos have been investigated from different
approaches \cite{Robnik84,BerryRobnik,Tomsovic94,Jacquod97} but there is not
an universal characterization up to now.

Most of the initial impulse for the development of Random Matrix Theory came
from nuclear physics. Wigner was the first to think of nuclear interactions
from a statistical point of view, renouncing to the exact knowledge of the
system and trying to analyze generic spectral properties instead
\cite{Wigner67}. The main difference with ordinary statistical mechanics is
that one renounces not to the exact knowledge of the state of the system
but to the nature of the system itself, the nature of the interaction, and
thus averages are calculated not with an ensemble of systems but with an
ensemble of hamiltonians: these are the random matrices. The first experimental
verification was carried out on the so-called Nuclear Data Ensemble (NDE),
a set of about 1700 data on proton and neutron resonances above the one-nucleon
emission threshold, the agreement with RMT being excellent \cite{Haq82}. As
nuclei are invariant under time reversal, the matrix representation of the
Hamiltonian can accordingly be chosen real and symmetric and thus the
Gaussian Orthogonal Ensemble (GOE) is the one to be used in this case.
Hence, one can say that in the high-energy region the picture is clear and one
can safely state that nuclei are chaotic systems. In the low-energy domain
the situation is less clear because of poorer statistics and uncertainties
in the experimental spectra. However, much effort has been dedicated to
analyze the experimental data and shed light on this issue. There is not
a general result but the type of energy level fluctuations depends on the
nuclear mass region and several factors. For example, for light and
spherical nuclei they are close to GOE, but for collective states in deformed
nuclei they are closer to Poisson, and in other cases the situation is
intermediate. For a complete review on the issue see \cite{Gomez11}.

This kind of statistical analysis was applied to the hadron mass spectrum in
\cite{Pascalutsa03} obtaining a chaotic-like behavior.
In \cite{Fernandez07} the spectral-statistic techniques have been used to
compare the experimental baryon spectrum with theoretical ones, focusing on
the problem of missing resonances. The main result of this work is that the
spectral fluctuation properties of theoretical quark-model spectra are
incompatible with those of the experimental spectrum, being the experimental
closer to GOE while the theoretical incompatible with GOE and closer to
Poisson. Given that the lack of levels in a spectrum produces a lost of
correlations among levels and thus a displacement towards the Poisson
distribution \cite{Bohigas04}, it is the experimental spectrum the one which
should be more uncorrelated, that is, closer to Poisson, because of the lack
of the missing resonances with respect to the theoretical models, but in fact
the situation is just the opposite. Hence, quark models, as they are
presently built, lack a very relevant property of the experimental spectrum:
its chaotic behavior. Thus, they may not be suitable to reproduce the low-lying
baryon spectrum, and, therefore, to predict the existence of missing resonances.
In \cite{Munoz12} this work has been extended to the meson spectrum,
employing an improved version of the approach used in \cite{Fernandez07}.
Also for mesons, the fluctuation properties of the experimental spectrum
are closer to GOE predictions, safely incompatible with Poisson, with an
estimation of 78\% of chaos.
Moreover, it is also tested that the analysis is robust against the inclusion
of the error bars associated to the experimental data.
For the theoretical models in this case, five of the six which have been
analyzed, including the lattice QCD calculation by the Hadron Spectrum
Collaboration (HSC) at JLab, are incompatible with chaos and closer to Poisson,
as for baryons. Only one of the quark models predicts an intermediate spectrum
with an estimation of 63\% of chaos. Thus, all the theoretical models but one
predict spectra with fluctuation properties incompatible with the experimental
one. This is especially shocking for the lattice QCD spectrum, as lattice QCD
is currently the only tool available to compute low-energy observables
employing QCD directly. Thus, the current state-of-the-art calculation does
not describe properly the statistical properties of the meson spectrum.

With this paper, we aim to fill some gaps coming from the aforementioned
works. First, we give a complete description of all the statistical tools
to deal with the kind of spectra present in the low-lying regions of
few-particle interacting quantum systems, with special emphasis on the new
method to perform a proper analysis, taking into account the shortness of
the spectral sequences. Thus, besides the meson and baryon
mass sequences, it can be also applied to other quantum spectra which
presents this problem, like for example the atomic nucleus. Second,
we perform the new analysis used for mesons in \cite{Munoz12} on the baryon
spectrum, refining and
updating the conclusion obtained in \cite{Fernandez07}. For both,
baryons and mesons, we perform the analysis on the last updated experimental
data from the Review of Particle Physics \cite{PDG2014}, and compare to
theoretical models,
giving major support to the conclusions previously obtained.

The article is organized as follows: In section \ref{sec:analysis} the
techniques used in the analysis of the spectra (experimental and
theoretical) are described. In section \ref{sec:results} we present the
results for the experimental baryon and meson spectra with comparison
to theoretical models. And in section \ref{sec:conclusions} we summarize
the results and state the main conclusions of the analysis.

\section{Statistical analysis}\label{sec:analysis}

Prior to any statistical analysis of the spectral fluctuations one has to
accomplish some preliminary tasks. First of all, it is necessary to take into
account all the symmetries that properly characterize the system. It is well
known that mixing different symmetries deflects the statistical properties
towards the Poisson statistics \cite{Molina06}. Hence, it is necessary to
separate the whole spectrum into sequences of energy levels involving the same
symmetries, that is, values of the good quantum numbers. The usual symmetries
associated to baryons are spin ($J$), isospin ($I$), parity ($P$), and
strangeness; and for mesons the same ones plus $C-$parity ($C$). Strangeness
can be dropped due to the assumption of flavor $SU(3)$ invariance. Therefore,
the baryon spectrum is split into sequences with fixed values of $J$, $I$ and
$P$, and the meson spectrum with fixed values of $J$, $I$, $P$ and $C$.

The energy spectrum of a quantum system is fully characterized by its
level density $g(E)$. It can be split into a smooth part
$\overline{g}(E)$, giving the secular behavior with the energy, and
a fluctuating part $\widetilde{g}(E)$, which is responsible for the
statistical properties of the spectrum to be analyzed \cite{Gutz71}. Thus, the 
fluctuation amplitudes of the latter are modulated by $\overline{g}(E)$
and, therefore, in order to compare the statistical properties of
different systems or different parts of the same spectrum the main trend
defined by $\overline{g}(E)$ must be removed.
The standard procedure by which it is removed is called the \textit{unfolding}.
It consists in locally mapping the real spectrum $\{E_i\}_{i=1,\ldots,N}$ into
another one $\{\varepsilon_i\}_{i=1,\ldots,N}$ with constant mean level density.
This can be done by means of the following transformation:
\begin{equation}
\varepsilon_i = \overline{m}(E_i), \;\;\; i=1,\ldots,N
\end{equation}
where $\overline{m}(E)$ is the smooth part of the cumulated level density
$m(E)$, which counts the number of levels whose energy is equal or less
than $E$, and as the level density $g(E)$, can also be separated
into a smooth part and a fluctuating part, and $N$ is the dimension of the
spectrum. The transformed level density $\rho(\varepsilon)$ in the new energy
variable $\varepsilon$ is such that $\overline{\rho}(\varepsilon)=1$, as
required. This general method of unfolding is called \textit{global unfolding}
to distinguish it from the \textit{local unfolding}, which we describe
in the next paragraph. In practical cases, the unfolding procedure can be
a difficult task for systems where there is no analytical expression for the
mean level density $\overline{g}(E)$, and it must be stressed that a correct
choice of $\overline{g}(E)$ is very important, as if it is not accurate enough
it will introduce errors in the fluctuation measures spoiling the statistical
analysis \cite{Molina02}.

When there is no natural choice for $\overline{g}(E)$ one can resort to simple
methods like the \textit{local unfolding}, in which this function is assumed
to be approximately constant in a window of $v$ levels on each side of a given
energy level $E_k$, and is given by
\begin{equation}
\overline{g}(E_k) = \frac{2v}{E_{k-v}-E_{k+v}}.
\end{equation}
It must be noted that this procedure can only be used to study short range
correlations, and it fails to account for the long range correlations of
the spectrum spoiling the relationship between the spectral fluctuations
and the regular or chaotic regime of the system \cite{Molina02}.

Since the experimental baryon and meson spectra have been divided in
very short sequences of levels, we will use the local unfolding procedure.
First of all, we can consider that the variation of $\overline{g}(E)$
along these sequences is negligible, since they are short enough.
Second, as it is not possible to study long-range correlations for
these sequences, the main disadvantage of the local unfolding procedure
does not apply to this case. It is important to remark that the usual
measures for short-range correlations, like the nearest neighbor spacing
distribution which we use in this work, are not usually spoiled by local
unfolding techniques.

The procedure we use in this paper is as follows.
Let $\{ E_i,\;i=1,2,\dots,l_x \}_X$ be an energy-level sequence
characterized by the set $X$ of good quantum numbers, and the
distances between consecutive levels, \mbox{$S_i = E_{i+1} - E_i$}. Thus,
assuming that the mean level density is constant along the sequence,
we can calculate the average value of the spacing between consecutive levels
$\left<S \right> = 1/\overline{g}(E) =
\left(l_x-1\right)^{-1}\sum_{i=1}^{l_x-1} S_i$ and use it to rescale
the level spacings to obtain the quantities $s_i =S_i / \left<S \right>$,
called generically nearest neighbor spacings (NNS). For the rescaled
spectrum the mean level density $\overline{\rho}(E)= 1$, and
$\left<s \right>=1$, thus the unfolding is performed. \vskip2ex

In this paper, the statistical properties of the NNS are studied by
means of the nearest neighbor spacing distribution
(NNSD) \cite{Mehta04}, denoted $P(s)$, which gives the number of spacings
lying between $s$ and $s+ds$, normalized to 1, that is, the probability
that the spacing between two consecutive unfolded levels lies between
$s$ and $s+ds$. The NNSD follows the Poisson distribution for generic
integrable systems \cite{Berry77}:
\begin{equation}
P_P(s) = \exp(-s)
\end{equation}
while chaotic systems with time reversal and rotational invariance are well
described by the GOE of random matrices, whose NNSD follows the Wigner surmise
\cite{Bohigas84}:
\begin{equation}
P_W(s) = \frac{\pi s}{2} \exp \left(-\frac{\pi s^2}{4} \right)
\end{equation}
For intermediate cases between integrable and chaotic systems, one of the most
frequently used is the Berry-Robnik distribution \cite{BerryRobnik}, based
on the principle of uniform semiclassical condensation (PUSC) \cite{Robnik84},
which states that certain spectral characteristics can be understood
by accounting for the separate chaotic and integrable regions in phase space.
Then, denoting by $f$ the volume fraction of the regular phase space,
the Berry-Robnik distribution is written as
\begin{widetext}
\begin{equation}
P_{BR}(f,s) = \left[f^2 \text{erfc} \left( \frac{\sqrt{\pi}(1-f)}{2}s \right) + \left( 2f(1-f)
+ \frac{\pi}{2}(1-f)^3s\right)\exp \left( -\frac{\pi}{4}(1-f)^2s^2\right)\right]\exp(-s).
\end{equation}
\end{widetext}

However, despite local unfolding is the usual way to deal with short sequences,
it is important to point out that it may cause a distortion on the actual
$P(s)$, preventing a direct comparison with the theoretical predictions
(Wigner or Poisson). This is a key point in this work, as we have to analyze
spectra which have to be split in very short sequences and, as we will show,
the effect is quite important to ignore it and do just the usual comparison
to theoretical predictions. Inasmuch as $\left<s \right>=1$
for every spacing sequence no one of the spacings can be greater than $l-1$,
where $l$ is the sequence length, and therefore the $P(s)$ distribution must
exhibit a sharp cutoff at $s=l-1$. When $l$ is large enough this cutoff
is irrelevant due to the exponential and Gaussian decays of the Poisson and
Wigner distributions. But obviously, this is not the case for smaller values
of $l$.

\begin{figure*}
\begin{center}
\rotatebox{0}{\scalebox{0.5}[0.4]{\includegraphics{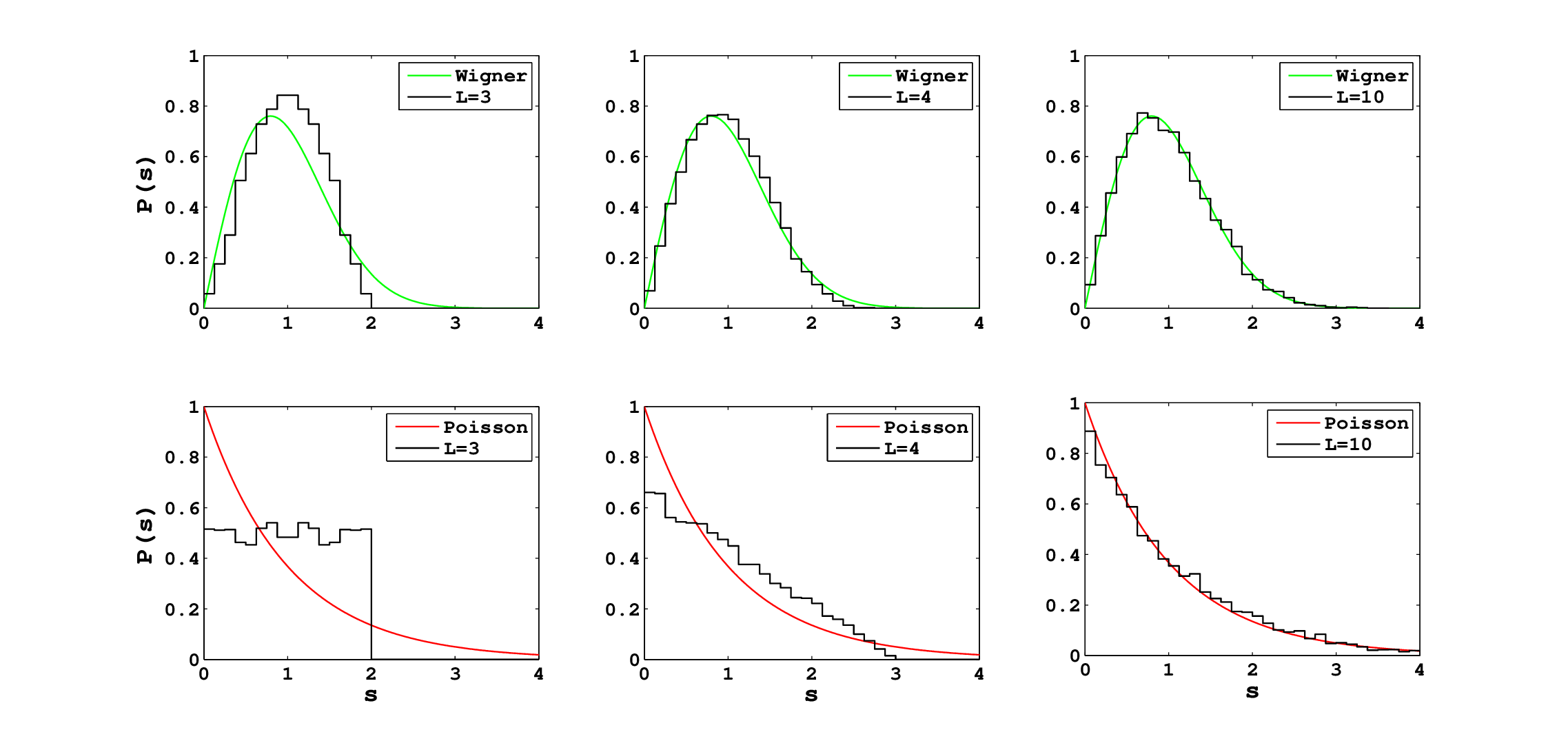}}}
\end{center}
\caption{(color online.) Nearest neighbor spacing distribution for GOE (up) and Poisson (down)
spectra (histograms) divided in sequences of lengths $l=3,4,10$ levels in which
a local unfolding has been performed, compared to the theoretical predictions,
the Wigner surmise and the Poisson distribution (solid lines).}
\label{Ejemplo_distorted}
\end{figure*}

In Fig. \ref{Ejemplo_distorted} we show some examples which illustrate
the problem clearly. We have performed the local unfolding on GOE and Poisson
spectra which have been divided in sequences of length $l=3, 4$ and 10 levels.
Each panel correspond to an ensemble of 100 spectra with about 100 levels
each, that is, similar to the dimensions of the spectra analyzed in this work.
The cutoff at $s=2$ and 3 can be clearly observed in the first two panels
for GOE (up) and Poisson spectra (down), and how the shape of the $P(s)$
distribution changes with respect to the theoretical predictions.
This misleading effect is specially relevant for Poisson sequences:
GOE-like ones are less affected due to the Gaussian decay of the tail of
the distribution. Already for $l=10$ one can say the effect is negligible.
However, we have
to deal with many short sequences when analyzing the low-lying spectra
of baryons and mesons, and therefore the correct treatment of this
difficulty is a key point to infer well supported conclusions.

This problem was taken into account in \cite{Fernandez07} by building
GOE-like and Poisson-like spectra distorted in the same way by the unfolding
procedure as the spectrum which is being studied in each case.
That is, by dividing the GOE and Poisson spectra in the same number of
sequences with the same lengths as the spectrum under study and performing
a local unfolding in the same way. These distributions built {\it ad hoc}
for the spectrum which is being studied are thus more adequate as reference
distributions for comparison than the theoretical predictions.

In this work we take a step further. Instead of building just one GOE-like
and one Poisson-like reference distorted spectra to compare, we generate
an ensemble of 1000 realizations, and their average will play the role of
{\it theoretical distributions} for comparison with the data in each case.
In this way we get a much smoother distribution to compare than with only
one sample, which could be too small to be representative of the
corresponding theoretical distribution.
We will denote these {\it distorted theoretical predictions} as $P_{DW}(s)$
for the distorted Wigner distribution, $P_{DP}(s)$ for the distorted Poisson
and $P_{DBR}(s)$ for the distorted Berry-Robnik in the cases when it is
necessary to build also an intermediate distribution.

In order to have a quantitative comparison of the data with the
reference distributions we shall use the Kolmogorov-Smirnov (K-S) test
\cite{KS}, which compares two samples in order to decide if the null
hypothesis that both belong to the same distribution can be rejected or not.
The statistic $D$ calculated in this test is the largest absolute deviation
between the two sample cumulative distribution functions. And the
obtained $p$-value, which can be used to evaluate the result of the
test, corresponds to the tail probability associated with the observed value
of $D$, that is, the probability, under the null hypothesis, of obtaining
a value of the test statistic $D$ as extreme as that observed.
The usual limit to reject the null hypothesis is $p \lesssim 0.10$. Thus,
much larger $p$-values do not allow to reject the null hypothesis and
much smaller $p$-values allow to safely reject it.

Complementary information can be gained by calculating the moments of the NNS
distributions, as they are univocally determined while the distribution itself
is sensitive to the bin size. Gathering together all the spacings $s_i$, the
$k$-th moment, $M^{(k)}$ is calculated as $M^{(k)} = (d-n)^{-1}
\sum_{i=1}^{d-n} s_i^k$, where $d$ stands for the spectrum dimension and
$n$ is the number of spacing sequences.

Finally, we perform a test in order to check if our analysis is robust against
the inclusion of the error bars associated to the experimental data.

\section{Results}\label{sec:results}

In this section we present the results of the statistical analysis, first
for baryons and second for mesons. In each case we show first the analysis
of the experimental spectrum and then that of the spectra obtained from
theoretical models.

\subsection{Baryons}

\subsubsection{Experimental spectrum}\label{sssec:barexp}

We have taken all the resonance states from the Review of Particle Physics
(RPP) \cite{PDG2014} up to 2.2 GeV. After splitting the
spectrum in sequences with the same $J$, $I$ and $P$, we have 53 levels
distributed in 14 sequences (only sequences with more than two levels are
considered).

\begin{figure}
\begin{center}
\rotatebox{0}{\scalebox{0.5}[0.5]{\includegraphics{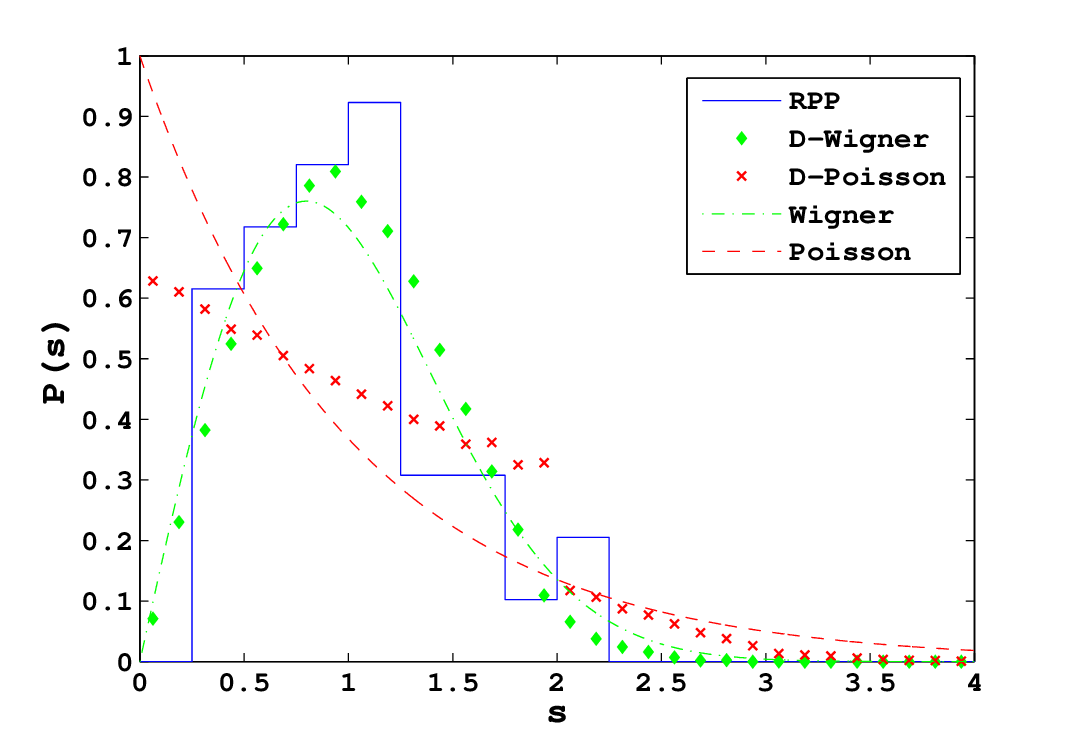}}}
\end{center}
\caption{(color online.) NNSD for the experimental baryon spectrum from the RPP
(histogram) compared to the distorted reference distributions (Wigner, $P_{DW}(s)$ (diamonds) and Poisson, $P_{DP}(s)$ (crosses)) and to the theoretical distributions
(Wigner surmise (dash-dotted) and the Poisson distribution (dashed)).}
\label{NNSDbar1}
\end{figure}

Fig. \ref{NNSDbar1} shows the $P(s)$ distribution for the experimental
spectrum compared to the Wigner surmise $P_W(s)$, the Poisson
distribution $P_P(s)$ and the corresponding distorted $P_{DW}(s)$ and
$P_{DP}(s)$, which are more adequate to compare, as it has been explained in
the previous section. It can be seen that the distortion is quite noticeable
in this case, especially for the Poisson distribution. A cut at $s=2$ can be
observed, as it is expected, because most of the sequences in which the
spectrum is divided in this case contain only three levels. To the naked eye,
the experimental $P(s)$ seems closer to the Wigner distribution than to the
Poisson one.
Its most relevant signature is the behavior at small spacings. As it is clearly
shown in the figure, $P(s) \xrightarrow[s \to 0]{} 0$. This feature is called
``level repulsion'' and it is a trademark of chaotic (Wigner-like) spectra,
whereas for Poisson sequences $P(0) \neq 0$. Moreover, a quantitative measure
is needed before obtaining a conclusion. To do so, we perform
the K-S test with the null hypothesis that the experimental distribution
coincides with the reference distribution $P_{DW}(s)$ or $P_{DP}(s)$ against
the hypothesis that both distributions are different. The results for the
$p$-value obtained in each case are
$p_{DW}=0.82$ and $p_{DP}=0.26$, that is, though the distribution seems
closer to Wigner, none of the null hypotheses can be rejected (the usual
limit for the $p$-value is $p \lesssim 0.10$). Thus in this case one could
try to compare the $P(s)$ with a Berry-Robnik distribution in order
to asses how close the fluctuations are to one limit or the other. But in
fact, we find that there is no Berry-Robnik distribution $P_{DBR}(f,s)$ which
fits the experimental $P(s)$ better than the Wigner distribution $P_{DW}(s)$
itself. This is clearly due to the strong repulsion shown by the experimental
$P(s)$, as pointed out above.

\begin{figure}
\begin{center}
\rotatebox{0}{\scalebox{0.5}[0.5]{\includegraphics{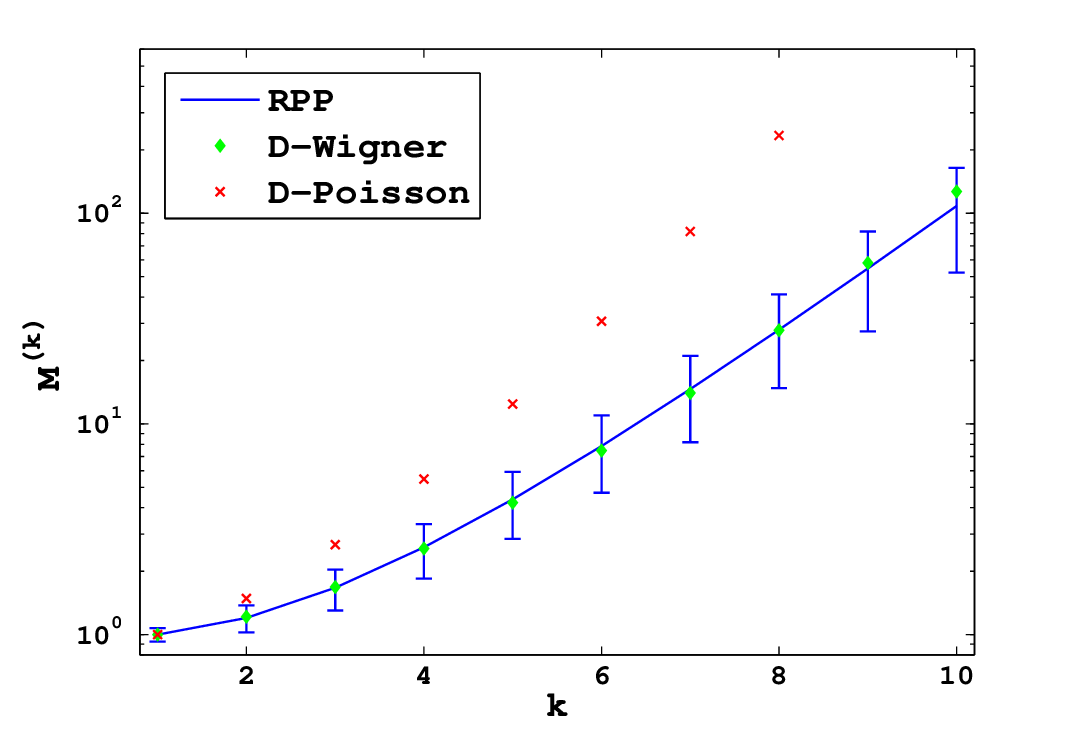}}}
\end{center}
\caption{(color online.) Moments of the NNSD, $M^{(k)}$, 
for the experimental baryon spectrum (solid line with error bars), 
compared to those of the distorted distributions: 
Wigner (diamonds) and Poisson (crosses). }
\label{moments_expbar}
\end{figure}

Fig. \ref{moments_expbar} displays the moments $M^{(k)}\!, \; 1 \le k \le 10$
for the experimental spacing distribution (the error bars correspond to the
standard deviation), as well as the $M^{(k)}$ corresponding to
the distorted distributions $P_{DW}(s)$ and $P_{DP}(s)$. It is shown that the
moments of the distorted Poisson distribution are outside and far away from
the error bars. Although the moments of $P_{DW}(s)$ are compatible with the
experimental data.

\begin{figure}
\begin{center}
\rotatebox{0}{\scalebox{0.5}[0.5]{\includegraphics{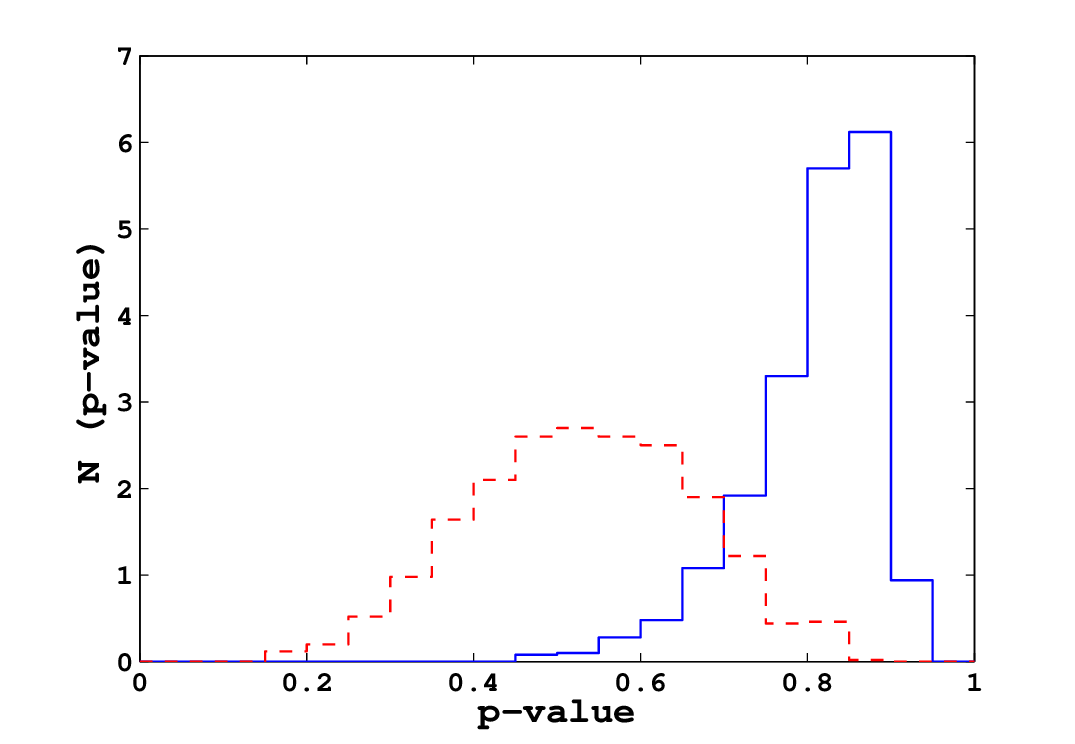}}}
\end{center}
\caption{(color online.) Distributions of the $p$-values of the K-S test for the 1000
\textquotedblleft realizations\textquotedblright \: of the experimental baryon
spectrum within the error bars, for the null hypothesis that the distribution
coincides with Wigner (solid histogram) and with Poisson (dashed histogram).}
\label{KSbar}
\end{figure}

Finally, it remains to test if our analysis is robust against the inclusion
of the error bars associated to the experimental data.
In order to see what happens if we do not consider the experimental masses
as exact but randomly variable inside the interval given by the error bars,
we will consider the experimental energies as Gaussian random variables with
mean equal to the RPP estimation and variance equal to
the corresponding error bar, we generate 1000 ``realizations'' of the
experimental spectrum and analyze them in the same way as the original one.
First, we build
the NNSD and perform the K-S test for each of them. Fig. \ref{KSbar} shows
the histograms of the resulting $p$-values for the comparison to $P_{DW}(s)$
and $P_{DP}(s)$.
The distribution of $p_{DW}$-values is narrowly concentrated
in the region of high values with $\left<p_{DW}\right> = 0.81 \pm 0.08$, that
is, it remains practically unchanged with respect to the $p_{DW}$ obtained for
the original experimental spectrum, thus indicating that the result is robust.
The distribution of $p_{DP}$-values is more spread and the mean value
$\left<p_{DP}\right> = 0.53 \pm 0.13$, which is higher than the one obtained
for the original spectrum, but still reasonable because if the energy levels
are allowed to fluctuate independently (in this case the fluctuation is induced
by the error bars) the correlations are usually weakened and thus the
statistics can be displaced towards Poisson.

To summarize, our analysis is fairly robust against experimental errors
and allows us to conclude that the statistical properties of the experimental
spectrum are much closer to the Wigner than to the Poisson limit, that is,
to chaos than to integrability, though none of the hypotheses can be safely
rejected. At this point one could think that the Poisson limit cannot be
completely discarded because of the missing resonances: if we suppose that
the statistical properties follow the Wigner prediction but there are missing
levels in the spectrum, then it is displaced towards the Poisson prediction
\cite{Bohigas04,Molina06}. If this were the case, then the spectra from
theoretical models, if they are complete, would be much closer to Wigner and
more clearly incompatible with Poisson. This point is analyzed in the next
section.

\subsubsection{Theoretical spectra}

Here we analyze the three theoretical spectra from quark models which
were analyzed in \cite{Fernandez07}, but now with comparison to these new
{\it theoretical predictions}, the distorted distributions. In this case
we do not expect the effect of distortion to be so noticeable as for the
experimental spectrum, as the dimensions of the sequences are not so small.
Table \ref{tableKSbar} displays relevant information on the experimental
and the three theoretical spectra: their dimension $d$, the number $n$
of pure sequences included in the analysis and the total number of spacings,
which is equal to $d-n$. We call CI the spectrum from the model by Capstick
and Isgur \cite{Capstick86}, which is a relativized quark model where the
interaction is built employing a one gluon exchange potential and confinement
is achieved through a spin-independent linear potential. It is the
immediate and essentially unique generalization of the model by Godfrey and
Isgur for mesons \cite{Godfrey85}, that is, from $\overline{q}q$ to $qqq$.
L1 and L2 are the spectra from the model by
L\"oring {\it et al.} \cite{Bonn01}, which is a relativistically covariant
quark model based on the three-fermion Bethe-Salpeter equation with
instantaneous two- and three-body forces (already used in \cite{Koll00}
for mesons).

\begin{figure}
\begin{center}
\rotatebox{0}{\scalebox{0.5}[0.7]{\includegraphics{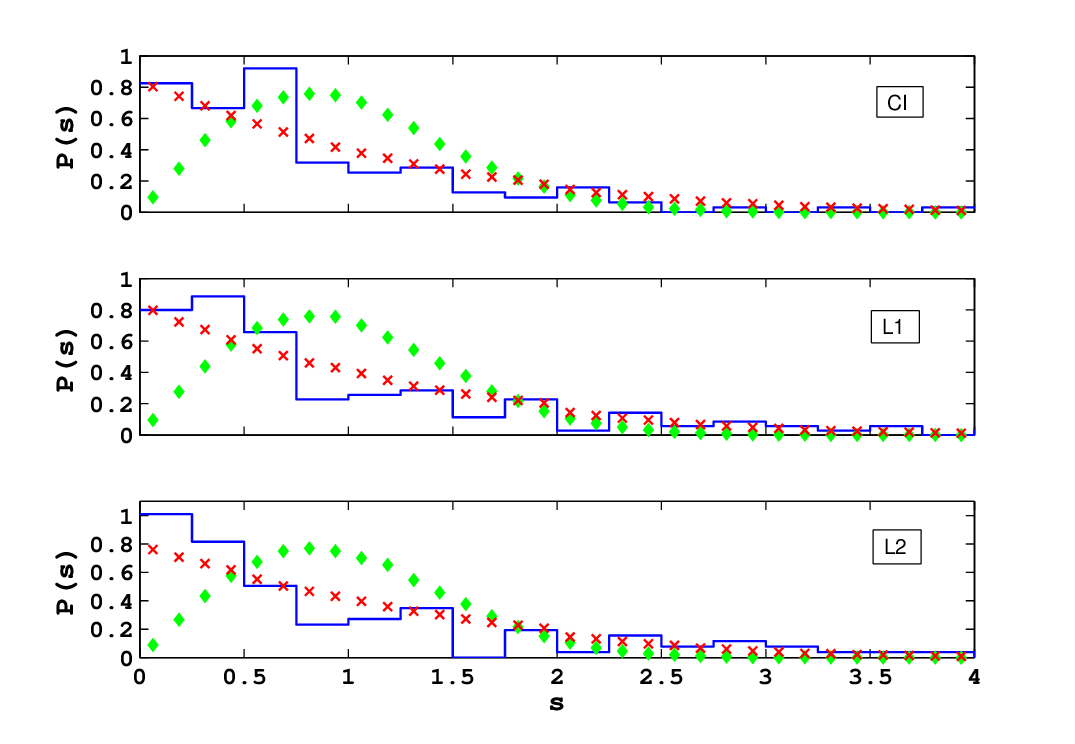}}}
\end{center}
\caption{(color online.) NNSD for the theoretical baryon spectra: i) Top,
set CI by Capstick and Isgur \cite{Capstick86}; ii) middle, set L1 by L\"oring
{\it et al.} \cite{Bonn01}; iii) bottom, set L2 by L\"oring {\it et al.}
\cite{Bonn01}; compared to the distorted distributions: Wigner, $P_{DW}(s)$
(diamonds) and Poisson, $P_{DP}(s)$ (crosses).}
\label{NNSDci}
\end{figure}

Fig. \ref{NNSDci} shows the NNSD for the three theoretical spectra,
compared to the distributions $P_{DW}(s)$ and $P_{DP}(s)$. As expected,
the distortion due to the unfolding is not so appreciable in this case, and
thus, the result of the analysis remains the same as in \cite{Fernandez07}.
The result is also confirmed by the K-S test. In Table \ref{tableKSbar} the
$p$-value of the K-S test for the experimental and the theoretical spectra
are shown. It is seen that all the theoretical spectra are incompatible with
the Wigner distribution ($p_{DW} \sim 10^{-4}$), whereas the experimental one
seems to be closer to the Wigner than to the Poisson distribution.

\begin{table}
\begin{center}
\caption{$p$-values of the K-S test for the experimental and
the theoretical baryon spectra. The null hypotheses are that the NNSD coincides
with the distorted Wigner surmise ($ p_{\sst DW}$)
and the distorted Poisson distribution ($ p_{\sst DP}$). $d$ stands for the
dimension of the spectrum and $n$ for the number of pure sequences.} \vspace*{0.3cm}
\begin{tabular}{c||c|c|c|c|c}
\hline
Set          &  $d$  &  $n$ &  $d-n$ &   $p_{\sst DW}$   & $ p_{\sst DP}$  \\ \hline
             &       &      &        &                 &               \\
RPP (data)   &  $53$ & $14$ & $39$   &      $0.82$     &    $0.26$     \\ 
CI           & $145$ & $19$ & $126$  & $5\cdot 10^{-4}$ &    $0.24$     \\
L1           & $164$ & $24$ & $140$  &     $10^{-4}$    &    $0.22$     \\
L2           & $124$ & $21$ & $103$  & $2\cdot 10^{-4}$ &    $0.20$     \\ \hline
\end{tabular}
\label{tableKSbar}
\end{center}
\end{table}

All these results confirm those obtained in \cite{Fernandez07}, giving major
support to the conclusions stated there. First, theory and experiment are
statistically incompatible. Second, the usual statement of missing resonances
cannot account for the discrepancies. As is well known, the existence of
missing levels in a spectrum deflects the statistical properties towards
Poisson \cite{Bohigas04,Molina06}. Thus, if the experimental spectrum is not
complete due to missing states, it should be closer to the Poisson distribution
than the theoretical ones. The situation is just the opposite. Hence, quark
models, as they are presently built, may not be suitable to reproduce the
low-lying baryon spectrum, and, therefore, to predict the existence of missing
resonances.

\subsection{Mesons}

\subsubsection{Experimental spectrum}

We have taken all the resonance states from RPP \cite{PDG2014} up to 2.5 GeV.
After splitting the spectrum in sequences with the same $J$, $I$, $P$ and $C$,
we have 129 levels distributed in 23 sequences.

\begin{figure}
\begin{center}
\rotatebox{0}{\scalebox{0.5}[0.5]{\includegraphics{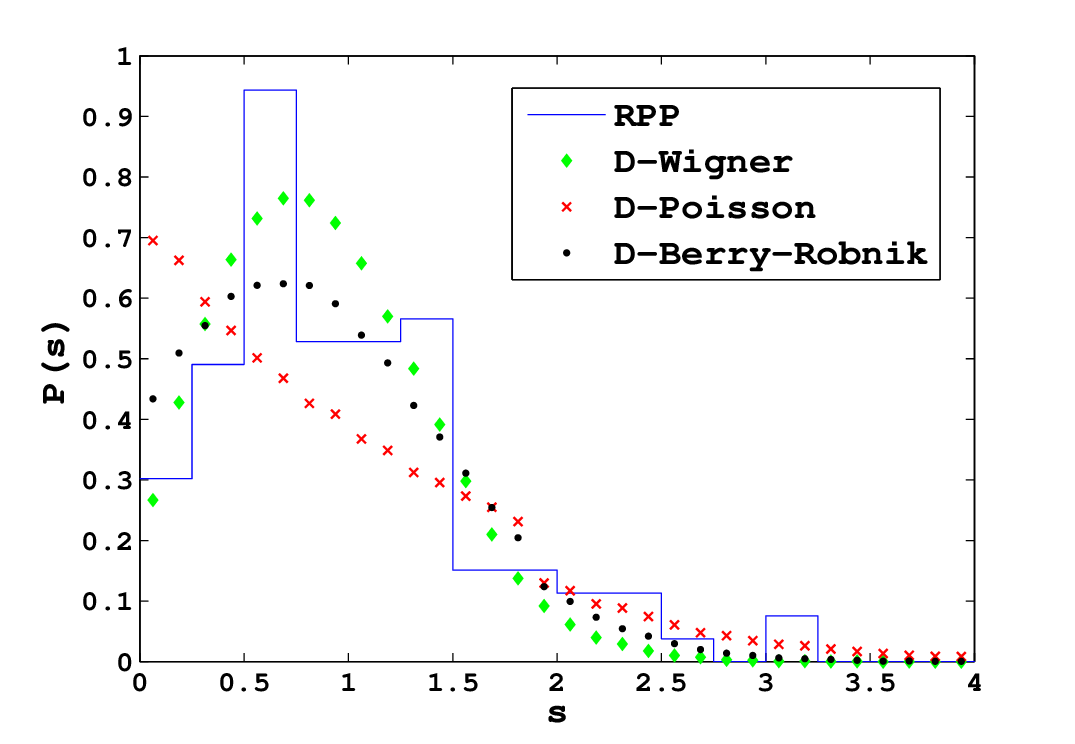}}}
\end{center}
\caption{(color online.) NNSD for the experimental meson spectrum from the RPP 
(histogram) compared to the distorted distributions: Wigner, $P_{DW}(s)$
(diamonds), Poisson, $P_{DP}(s)$ (crosses) and Berry-Robnik with $f=0.78$,
$P_{DBR}(0.78,s)$ (dots).}
\label{NNSDmes}
\end{figure}

Fig. \ref{NNSDmes} shows the $P(s)$ distribution for the experimental spectrum
together with the distributions $P_{DW}(s)$ and $P_{DP}(s)$. It seems that the
statistical properties of the experimental distribution are intermediate
between the Poisson and Wigner predictions, though closer to the latter.
Then, we fit the experimental $P(s)$ to a Berry-Robnik distribution
$P_{DBR}(f,s)$ and in this case, unlike for baryons, we do obtain a best fit
which is intermediate between Wigner ($f=1$) and Poisson ($f=0$), that is,
$f = 0.78 \pm 0.03$. Figure \ref{NNSDmes} also displays $P_{DBR}(0.78,s)$.

The results for the $p$-value from the K-S test for the comparison with
the three reference distributions are the following: $p_{DP}=0.13$,
$p_{DW}=0.38$ and $p_{DBR}=0.65$. That is, confirming what can be seen in
the figure, the statistical properties of the experimental meson spectrum
are intermediate between the Poisson and Wigner limits, though they are
closer to the latter since a Berry-Robnik distribution with $f=0.78$ fits
well the experimental NNSD. It is also worth to note that
$p_{DP}=0.13$ is close to the usual limit for the null hypothesis to be
rejected ($p \lesssim 0.10$).

\begin{figure}
\begin{center}
\rotatebox{0}{\scalebox{0.5}[0.5]{\includegraphics{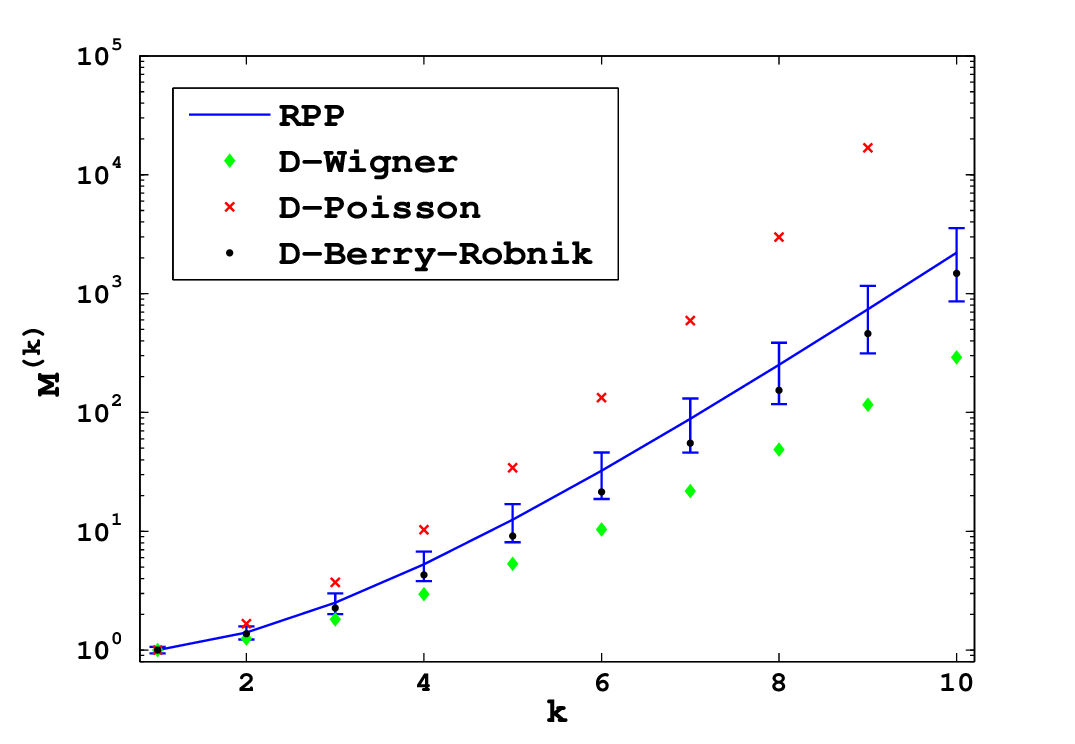}}}
\end{center}
\caption{(color online.) Moments of the NNSD, $M^{(k)}$, 
for the experimental meson spectrum (solid line with error bars), 
compared to those of the distorted distributions: 
Wigner (diamonds), Poisson (crosses)
and Berry-Robnik with $f=0.78$ (dots). }
\label{moments_expmes}
\end{figure}

Figure \ref{moments_expmes} displays the moments $M^{(k)}$ for the experimental
spacing distribution (the error bars correspond to the standard deviation),
and those corresponding to the distributions $P_{DP}(s)$, $P_{DW}(s)$ and
$P_{DBR}(s)$. It is shown that the moments of the distorted Poisson distribution
are outside and far away from the error bars. Those of $P_{DW}(s)$ are nearer
(note the logarithmic scale). And only the moments of $P_{DBR}(f,s)$ with
$f=0.78$ match the experimental result supporting our choice of $f$.

\begin{figure}
\begin{center}
\rotatebox{0}{\scalebox{0.5}[0.5]{\includegraphics{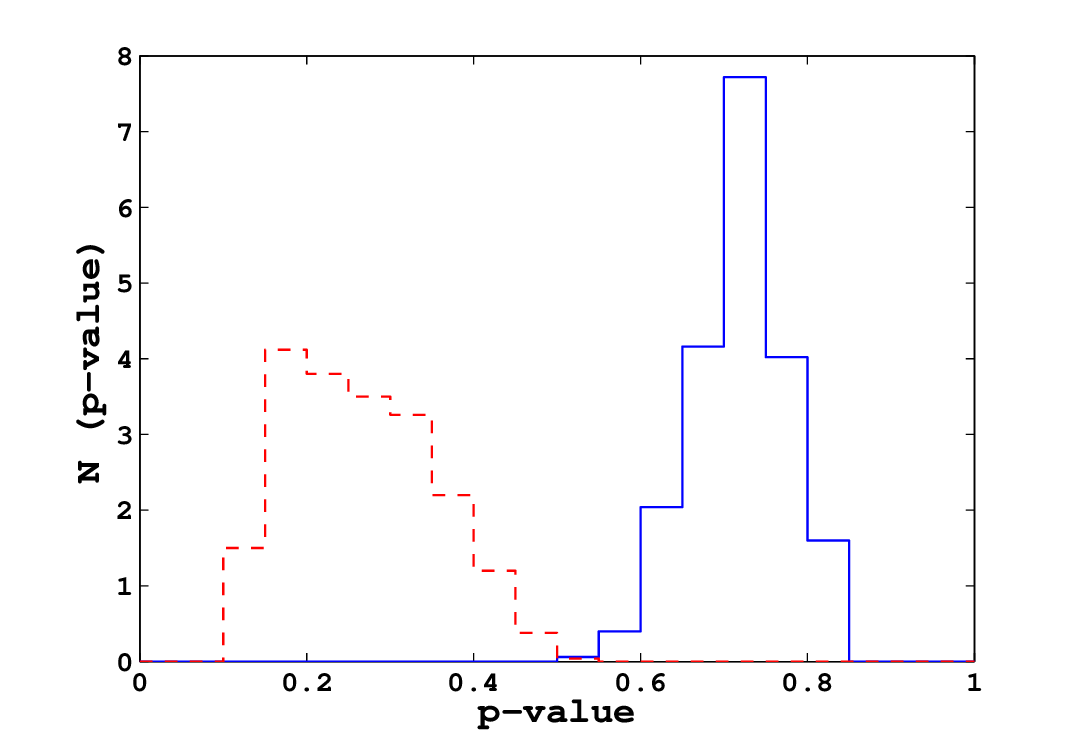}}}
\end{center}
\caption{(color online.) Distributions of the $p$-values of the K-S test for the 1000
\textquotedblleft realizations\textquotedblright \: of the
experimental meson spectrum within the error bars, for the null hypothesis that the distribution
coincides with Poisson (dashed histogram) and with Berry-Robnik for $f=0.78$ (solid histogram).}
\label{KSmes}
\end{figure}

Finally, we test if the analysis is robust against the inclusion of the error
bars associated to the experimental data. As for baryons, we generate 1000
``realizations'' of the experimental spectrum considering the experimental
energies as Gaussian random variables with mean equal to the RPP estimation
and variance equal to the corresponding error bar. In this case, we compare
the random realizations of the experimental spectrum with $P_{DP}(s)$
and with $P_{DBR}(0.78,s)$, as we have seen that it is the distribution which
better fits the experimental one. Figure \ref{KSmes} shows that the histograms
of the resulting $p$-values are separated with almost no
overlap. The distribution of $p_{DBR}$-values is concentrated in the
upper half with $\left<p_{\sst DBR}\right> = 0.72 \pm 0.06$, and the
histogram of the $p_{DP}$-values lies in the lower half with
centroid $\left<p_{\sst DP}\right> = 0.27 \pm 0.09$.  It is
important to  notice that for almost every 
\textquotedblleft realization\textquotedblright \: of the
experimental spectrum $p_{\sst DBR} > p_{\sst DP}$, 
sustaining the good agreement of the experiment with the
Berry-Robnik distribution 
for $f=0.78$. For the sake of completeness we have also
used as reference distribution the Wigner surmise, obtaining
$\left<p_{\sst DW}\right> = 0.44 \pm 0.11$.

To summarize, our analysis is fairly robust against experimental errors
and allows us to conclude that the
statistical properties of the experimental spectrum are intermediate
between the Wigner and Poisson limits, closer to the former and safely
incompatible with the latter. That is, mesons are much closer to chaotic
systems that to integrable ones. Moreover, a Berry-Robnik distribution
with 78$\%$ of chaos provides the best description of
the experimental NNSD.

Moreover, it is noteworthy to mention that a higher percentage of chaos is
inferred from the statistical analysis of the baryon spectrum. This result
is physically reasonable, as the three-quark system is more complex than the
two-quark one.

\subsubsection{Theoretical spectra}\label{ssec:theory}

Next we analyze six theoretical calculations of the light meson
spectrum and compare them to the results from previous section. These
are: (i) The classic model by Godfrey and Isgur (set GI)
\cite{Godfrey85}, which is a relativized quark model where the
interaction is built employing a one gluon exchange potential and
confinement is achieved through a spin-independent linear potential;
(ii) and (iii) are the fully relativistic quark models by Koll
\textit{et al.} (sets K1 and K2 which correspond, respectively, to
models $\cal{A}$ and ${\cal B}$ in \cite{Koll00}) based on the
Bethe-Salpeter equation in its instantaneous approximation, a flavor
dependent two-body interaction and spin-dependent confinement force,
being the last the difference between the two models; (iv) the
relativistic quark model by Ebert \textit{et al.} (set E)
\cite{Ebert09} based on a quasipotential (this calculation has the
disadvantage that isoscalar and isovector mesons composed by $u$ and
$d$ quarks are degenerate); (v) the effective quark model by Vijande
\textit{et al.} (set V) \cite{Vijande05}, based upon the effective
exchange of $\pi$, $\sigma$, $\eta$ and $K$ mesons between constituent
quarks; and (vi) the lattice QCD calculation by the Hadron Spectrum
Collaboration at JLab (set LQCD) \cite{Dudek11}.
Lattice QCD calculation does not include strange
mesons as the previous models, but it includes exotics such as the
isoscalar $J^{PC}=2^{+-}$ states, and it is computed at a high pion
mass of 396 MeV.

\begin{table}
\begin{center}
\caption[]{$p$-values of the K-S test for the experimental and the
theoretical meson spectra. The null hypotheses are
that the NNSD coincides with the distorted Wigner surmise ($ p_{\sst DW}$)
and the distorted Poisson distribution ($ p_{\sst DP}$). $d$ stands for the
dimension of the spectrum and $n$ for the number of pure sequences.} \vspace*{0.3cm}

\begin{tabular}{c||c|c|c|c|c}
\hline
Set  &    $d$  &  $n$  &  $d-n$   & $ p_{\sst DW}$ & $ p_{\sst DP}$ \\  \hline
     &       &      &        &              &               \\
RPP (data)  & $129$ & $23$ & $106$  &    $0.38$    &    $0.13$     \\ 
GI   &  $68$ & $17$ &  $51$  &    $0.84$    &    $0.41$     \\
K1   & $162$ & $38$ & $124$  &   $0.038$    &    $0.55$     \\
K2   & $162$ & $38$ & $124$  &   $0.005$    &    $0.43$     \\
E    & $190$ & $34$ & $156$  &   $0.083$    &    $0.21$     \\
V    &  $94$ & $18$ &  $76$  &    $0.51$    &    $0.56$     \\
LQCD &  $60$ & $15$ &  $45$  &   $0.033$    &    $0.44$     \\ \hline
\end{tabular}
\label{tableKS}
\end{center}
\end{table}

Table \ref{tableKS} displays relevant information on the six theoretical spectra, like their
dimension $d$, the number $n$ of pure sequences included in the analysis and the total number of
spacings, which is equal to $d-n$. It also provides the $p$-values obtained by applying the K-S test
to their NNSDs, taking as null hypotheses that the NNSD coincides either with $P_{DW}(s)$ or with
$P_{DP}(s)$. The first relevant outcome is that, according to the K-S test, the NNSDs
of sets K1, K2, E and LQCD are incompatible with the Wigner correlations and closer to the Poisson
statistics. Thus, the dynamics predicted by these models is essentially regular, while the
statistical properties of the experimental light meson spectrum show that the dynamical regime
should be chaotic.  This fact resembles the results obtained for baryons: while the fluctuations
of the experimental baryon spectrum are well reproduced by Wigner predictions, the theoretical
calculations give rise to spectra with Poisson statistics.

\begin{figure}
\begin{center}
\rotatebox{0}{\scalebox{0.5}[0.55]{\includegraphics{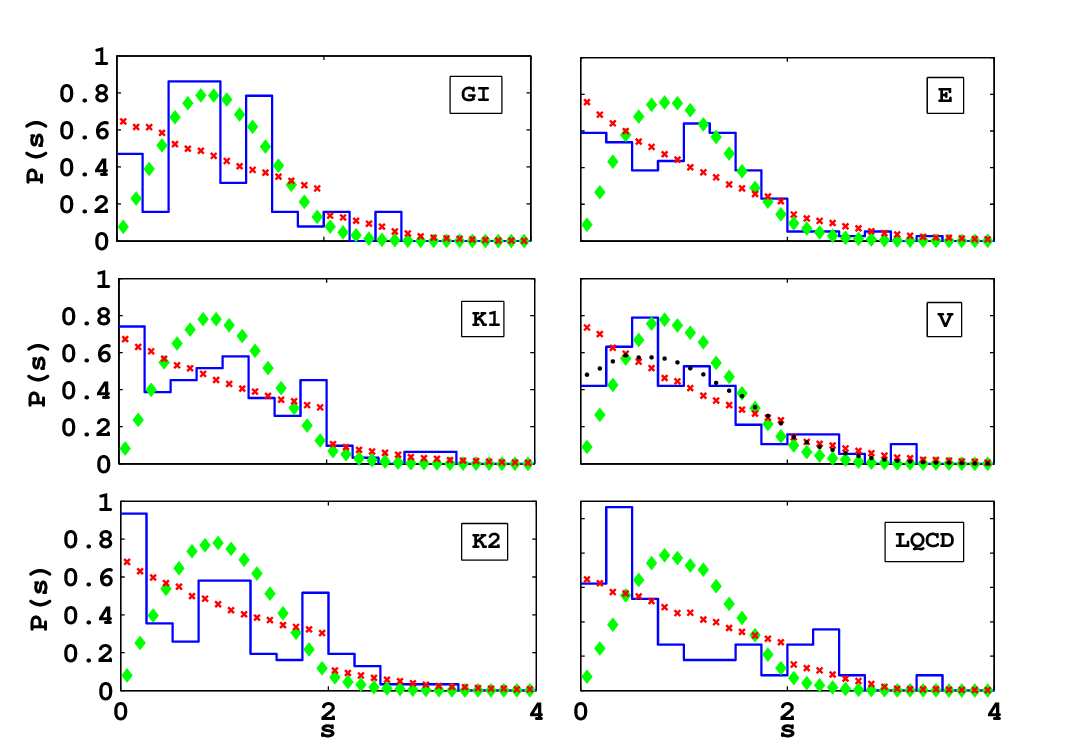}}}
\end{center}
\caption{(color online.) NNSDs for the theoretical meson spectra: 
i) Top left, set GI by Godfrey and Isgur \cite{Godfrey85}; 
ii) middle left, set K1 by Koll \textit{et al.} \cite{Koll00};
iii) bottom left, set K2 by Koll \textit{et al.} \cite{Koll00}; 
iv) top right, set E by Ebert \textit{et al.} \cite{Ebert09};
v) middle right, set V by Vijande \textit{et al.} \cite{Vijande05} 
and \textit{ad hoc} distorted Berry-Robnik with $f=0.63$; 
vi) bottom right, set LQCD by Dudek \textit{et al.} \cite{Dudek11}.
Distorted Wigner, $P_{DW}(s)$ is represented with diamonds, 
distorted Poisson, $P_{DP}(s)$ with crosses, and distorted Berry-Robnik,
$P_{DBR}(f,s)$ with dots.}
\label{NNSDteor}
\end{figure}

Figure \ref{NNSDteor} shows the NNSD of the six theoretical spectra. Sets K1 and E provide flat NNSDs with a cut at $s=2$.  The cut is expected as was explained in
section \ref{sec:analysis}.  When the Poisson distribution is distorted it flattens due to the
small amount of levels, so actually the NNSDs that we find for sets K1 and E are the ones we expect
from a Poisson distribution, confirming that these sets have less correlations than the experimental
data as the K-S test suggests.  The comparison between models K1 and K2 by Koll \textit{et al.} is
particularly interesting because they only differ on the confinement interaction and show how
important that interaction can be for the spectral statistics, hinting that it should be revised to obtain a better agreement with the experiment.

The result for set LQCD is particularly interesting because lattice QCD is currently the only tool
available to compute low-energy observables employing QCD directly.  We find that the current
state-of-the-art calculation in \cite{Dudek11} does not describe properly the statistical properties
of the meson spectrum. Lattice QCD NNSD is relatively close to the $P_{DP}(s)$ as it is
  shown in figure \ref{NNSDteor}.  This is evident at zero spacing where $P_{LQCD}(s=0) \approx
  0.6$, thus implying uncorrelated levels, as it was explained in section \ref{sssec:barexp}.  Our results
  remain unaltered if the statistical errors of the lattice QCD calculation are taken into account.
Thus, the LQCD calculation should be considered a step forward in lattice calculations but still far
away from being a description of the data or their structure.  It is not something unexpected given
that the LQCD set has been obtained at a pion mass of 396 MeV, far away from its physical mass, and
it is reasonable to expect a drastic change in the statistical properties when calculations get the
pion mass closer to its actual value. However, the fact that the lattice QCD calculation has a lot
less correlations than the experimental data, being practically uncorrelated, demands, besides the
need of bringing the calculation closer to the physical pion mass, a careful examination of the
approximations employed.

The results of the K-S test for sets GI and V are inconclusive because they
suggest that the models are compatible with both chaotic and integrable
dynamics. It is thus mandatory to take a close look
  to the NNSDs (figure \ref{NNSDteor}) before obtaining any conclusion.  Set GI NNSD has a strange
shape with peaks and dips, completely different from any of the usual distributions (Poisson, Wigner or Berry-Robnik), and therefore not close
at all to experiment (see figure \ref{NNSDmes}). It only has some similarity
with some very particular integrable systems whose $P(s)$ is equal to a
sum of $\delta$ functions, constituting an exception to the rule of Poisson
distribution \cite{Makino03}. On the contrary, set V displays a smooth NNSD,
which can be very well fitted to a distorted Berry-Robnik distribution with
$f=0.63 \pm 0.19$ (also displayed in figure \ref{NNSDteor}). Then we can
conclude that the model by Vijande {\em et al.} gives a better account of the
dynamical regime of the light meson spectrum.

\section{Conclusions}\label{sec:conclusions}

In this paper we have analyzed the spectral fluctuations of the experimental
and theoretical baryon and meson mass spectra in the context of Quantum Chaos.
Comparing the statistical properties of the spectra with Random Matrix Theory
(RMT) predictions is a tool to determine the dynamical regime of the system,
that is, whether the system is chaotic, regular or intermediate. We emphasize
that, besides the coincidence of the theoretical individual energies with
the experimental ones, the agreement in the statistical fluctuation properties
is also important, since they determine the dynamical regime and thus
can provide insight on the properties of the underlying interactions.

The statistical analysis is described stressing the fact that one has to be
very careful when dealing with spectra like those of the baryon and meson
masses, which must be divided in very short sequences to perform
the analysis, according to symmetry classes (sequences with the same quantum
numbers). We show a new method to take into account this problem.
It consists in building {\it distorted theoretical distributions} adapted to
compare with the spectrum under study. The distortion is induced by the local
unfolding procedure on the actual theoretical predictions exactly in the same
way as it is induced on the spectrum under study. For very short sequences,
the distorted theoretical predictions are more reliable than comparing directly
with the RMT predictions.

Once this analysis is carried out, we obtain that both experimental baryon
and meson spectra are closer to a chaotic behavior than to an integrable one.
The baryon spectrum seems to be more chaotic than the meson one, result that is
physically reasonable, as a 3-particle system is more complex than a 2-particle
one. The best description of the nearest-neighbor spacing distribution (NNSD)
for baryons is provided by the Wigner surmise, which is the prediction for
chaotic systems, whereas the best description of the NNSD for mesons is given
by the intermediate Berry-Robnik distribution with a 78\% of chaos.

We have also tested the robustness of the analysis against the inclusion of the
experimental errors, by performing the Kolmogorov-Smirnov (K-S) test on an
ensemble of spectra generated by considering the experimental masses as
Gaussian random variables with a mean given by the Review of Particle Physics
(RPP) estimation and variance equal to the corresponding error bar. In both
cases, baryon and meson mass spectra, the result shows that our analysis is
fairly robust against experimental errors.

As for the theoretical baryon spectra, the three spectra from quark models
which have been analyzed are clearly statistically incompatible with the
experimental one. From the K-S test, the hypothesis that the NNSD from quark
models coincides with the Wigner surmise can be rejected whereas the
experimental NNSD is clearly closer to the Wigner surmise than to the Poisson
one. Moreover, this discrepancy cannot be accounted for by the existence of the
missing resonances. It is well known that the lack of levels in a spectrum
deflects the statistical properties towards Poisson.
Thus, if this were the origin of the discrepancy, the experimental spectrum
should be closer to Poisson statistics than the theoretical ones, but the
situation is just the opposite. Hence, having present the importance of the
agreement in the statistical spectral properties and its relation with the
dynamical regime of the system, one can only state that quark models, as they
are presently built, may not be suitable to reproduce the low-lying spectrum,
and, therefore, to predict the existence of missing resonances.

In the case of mesons, of the six theoretical spectra which have been analyzed
only the one by Vijande {\it et al.} \cite{Vijande05} seems to be compatible
with the experimental one, with a NNSD well fitted with the intermediate
Berry-Robnik distribution with a 63\% of chaos. The other theoretical models,
including the Lattice QCD (LQCD) calculation, predict a regular or nearly
regular dynamics in clear contradiction with the experiment. The disagreement
with the LQCD spectrum is specially shocking as LQCD is currently the only tool
available to compute low energy observables employing QCD directly. Thus, we
find that the current state-of-the-art calculation in \cite{Dudek11} does not
describe properly the statistical properties of the meson spectrum. The failure
could be due to the fact that the calculation is made at an unrealistic pion
mass. For the quark models, further work is needed to study the origin of the
discrepancy with the experimental spectrum. In particular, it would be
interesting to study the differences of the model by Vijande {\it et al.}
with the other quark models, trying to find the signals of chaos.

\begin{acknowledgements}
The authors thank Dr. R. A. Molina and Dr. S. Melis for valuable comments.
This research has been conducted with support of the Spanish Ministry of
Science and Innovation grant FIS2012-35316.
\end{acknowledgements}

\end{document}